\documentclass[fleqn,usenatbib]{mnras}

\usepackage{newtxtext,newtxmath}
\usepackage[T1]{fontenc}

\DeclareRobustCommand{\VAN}[3]{#2}
\let\VANthebibliography\thebibliography
\def\thebibliography{\DeclareRobustCommand{\VAN}[3]{##3}\VANthebibliography}

\usepackage{graphicx}	
\usepackage{amsmath}	

\usepackage{amssymb}	
\usepackage{footnote}
\usepackage{hyperref}

\def\be{\begin{equation}}
\def\ee{\end{equation}}

\title[FAST discovery of long tidal tails in NGC 4490]{FAST discovery of long tidal tails in NGC 4490/85}

\author[Yao Liu et al.]{
Yao Liu, $^{1,3,4}$\thanks{liuyao@nao.cas.cn} 
Ming Zhu,$^{1,2,3,4}$\thanks{mz@nao.cas.cn} 
Haiyang Yu,$^{1,3,4}$
Mei Ai,$^{1,2,3,4}$
Peng Jiang,$^{1,2,4}$
Siqi Liu,$^{1,3}$
\newauthor
Ruilei Zhou$^{1,3,4}$
and Lixia Yuan$^{5}$
\\
$^{1}$ National Astronomical Observatories, Chinese Academy of Sciences, 20A Datun Road, Chaoyang District, Beijing 100101, China\\
$^{2}$Guizhou Radio Astronomical Observatory, Guizhou University Guiyang 550000, People's Republic of China\\
$^{3}$University of Chinese Academy of Sciences, Beijing 100049, China\\
$^{4}$CAS FAST key laboratory, NAOC, Chinese Academy of Sciences, Beijing 100101, China\\
$^{5}$ Purple Mountain Observatory and Key Laboratory of Radio Astronomy, Chinese Academy of Sciences, 10 Yuanhua Road, Qixia District, Nanjing 210033, China\\
}

\date{Accepted 2023 April 25. Received 2023 April 24; in original form 2023 February 3}

\pubyear{2023}

\begin{document}
\label{firstpage}
\pagerange{\pageref{firstpage}--\pageref{lastpage}}
\maketitle

\begin{abstract}
We report the discovery of a 100 kpc HI tail in the  merging 
galaxy pair NGC 4490/85 detected by the Five-Hundred-meter Aperture Spherical radio Telescope (FAST). 
The tidal tails extended in both the south and north directions, and they 
are much longer than that reported previously based on the VLA interferometric maps. 
The NGC 4490/85 is surrounded by a large gas envelope, and a starburst
low metallicity dwarf galaxy  MAPS 1231+42 is found to be connected 
with the gas envelope,  indicating that galaxy interaction trigged the
intense star formation in it. Based on the fact that the metallicity in MAPS 1231+42 is one order of magnitude lower
than that in the two disks of NGC 4490 and NGC 4485, we speculate 
that the gas near this galaxy should be primordial and could be due to 
gas inflow from the circum-galactic medium (CGM). We also found a collimated gas component pointing at a nearby dwarf galaxy KK 149,
suggesting that this galaxy might also  be interacting with the NGC 4490
pair. We discuss the possible origin of the long tidal tails and the extended gas envelope in this merging 
system based on the new data from FAST. 
\end{abstract}


\section{Introduction}
NGC 4485/90, also named Arp 269 (\citealt{arp1966atlas}), is a relatively
isolated low-mass galaxy pair which consist of a smaller irregular galaxy NGC 4485 and a larger barred spiral galaxy NGC 4490.  One remarkable feature of this system is that it is embedded in a very extended, low-density envelope of neutral hydrogen (HI).
Such a feature was first noted by \cite{1980A&A....91..341H} and
\cite{1980A&A....82..207V}. \cite{1998MNRAS.297.1015C} made a detailed HI study of this pair with the VLA, and found that the  HI envelope is elongated and extended for a total length of about 90 kpc, and it is approximately perpendicular to the
NGC 4490 disk. Based on this fact, they suggested  that the HI envelope can be explained by a galactic-scale bipolar outflow of HI driven by supernovae in NGC 4490.

Another interesting feature for the NGC 4485/90 pair is that they are considered to be an isolated analogue of the Magellanic Clouds (LMC and
SMC; e.g. \citealt{gardiner1996n}; \citealt{bekki2005formation}; \citealt{connors2006n}; \citealt{besla2010simulations}; \citealt{diaz2011constraining}; \citealt{besla2012role}; \citealt{guglielmo2014genetic}; \citealt{pardy2018models}). Both of them are good representative of 
closely interacting dwarf galaxies.
\cite{pearson2018modelling} modeled the NGC 4485/90 system  and they find 
that this pair's orbit appears to be an evolved version
of what the LMC-SMC might have looked like if they were not located near a massive galaxy, 
e.g. the Milky Way(MW). Both systems are prograde dwarf–dwarf interactions which can move a significant amount of gas to large distance. The LMC and SMC is 55 kpc to the Milky Way, but NGC 4485/90 is 310 kpc away (in projection) to its nearest massive neighbor galaxy NGC 4369.  \cite{pearson2018modelling}
also found a plausible encounter history that results in a baryonic mass distribution similar to the VLA HI map of \cite{1998MNRAS.297.1015C}. 
They shows that, when viewing from a special angle,  the tidal tail structure appears to be resemble a large extended gas envelope. Thus they suggest that the extended envelope is originated from tides alone, and not due to star formation outflows.

Recently, \cite{lawrence2020revealing} use a deep mid-infrared Spitzer image to show that NGC 4490 has a clear double nucleus structure, suggesting that NGC 4490 is a late-stage merger remnant. Thus the NGC 4485/90
system may have a longer and more complicated interaction history than previous thoughts.

In this work, we report the discovery of more tidal tail features in the NGC4485/90 pair which extend much longer than that detected with the VLA interferometer,
using the Five-hundred-meter spherical radio telescope (FAST).  This result
 shows that very long tidal tails 
are possible in isolated dwarf merging systems without the help of a massive halo(e.g. the LMC-SMC system).
We also detect HI emissions from two nearby dwarf galaxies which suggests that the interaction history in this system could be much more complicated than the 
simple two-body interaction model.

This paper is organized as follows: section 2 presents the observation and data reduction,
section 3 shows the results from FAST with emphasis on the new features found by FAST, and section 4 discusses the origin of the tail and envelope based on 
new features found.  Section 5 summarizes the conclusion.

 Based on the method of tip of red giant branch (TRGB), the distance is 8.8 Mpc for NGC 4485 and 6.5 Mpc for 
NGC 4490 (\cite{Sabbi2018}). Here we use an average distance of 7.14 Mpc for the pair in order to be able to 
directly compare with the simulation of \cite{pearson2018modelling}. All parameters can be scaled easily if a different distance is used.

\section{Observations and data reductions}
The extended tidal tail features in NGC 4485/90 region were first discovered in the FAST Extra-galactic HI Survey, which is a new survey for HI emission in the northern sky ($60 > DEC > -10$) over the
velocity range -2000 to 20000 km s$^{-1}$
(\citealt{2022RAA....22f5019K}; \citealt{Zhu2021}. FAST has a focal-plane 19-beam receiver system (\citealt{2019SCPMA..6259502J}) set in a hexagonal array and works in dual polarization
mode. It covers the frequency range from 1050 MHz to 1450 MHz. The FAST HI survey is carried out with the drift scan mode and the 19-beam receiver was rotated by 23.4 degrees so that the beam tracks are equally spaced in Declination with 1.1' spacing. The data presented here come from the drift scan observations conducted during Aug 09-27, 2021. We have also used the Multibeam on-the-fly (OTF) mode to map the NGC 4490 region to reveal more features of the tidal tails on Aug 15-16, 2021.  The OTF observations have almost the same setup as the drift scan observations, except that we used $10.3'$ scanning separation and $15"$ s$^{-1}$ scanning speed. 
The total on-source mapping time is about 4.5 hours.

For the backend, we choose the Spec(W) spectrometer which has 65536 channels covering the bandwidth
of 500 MHz for each polarization and beam, with a velocity spacing of 1.67 km s$^{-1}$ and a spectral resolution, after Hanning smoothing, of 5.1 km s$^{-1}$ which is sufficient for resolving fine spectral structures and obtaining accurate column densities as well as velocity fields.  The system temperature ranges between 18-22 K for all beam/polarization
channels. The half-power beam width (HPBW) was about $2.9'$ at 1.4 GHz for each beam. 

Flux calibration was performed by injecting a 10K calibration signal (CAL) every 32 seconds for a duration of 1 second to calibrate
the antenna temperature. The data were reduced using the HI Pipeline reduction software (which performed band pass calibration, flux calibration and RFI flagging), developed by \cite{Wang2023ApJ}. Baseline correction was performed using the asymmetrically reweighted penalized least squares algorithm (\citealt{2015Ana...140..250B}) which has been
successfully applied to various spectral analyse (e.g., \citealt{Zeng2021MNRAS.500.2969Z}; \citealt{Zhang2021RAA}). The RFI mitigation techniques are described in \citealt{Zhang2022RAA}).

Spectra were recorded every second, yielding approximately 38 samples 
per beam in the Right Ascension direction, while in the DEC direction the spacing is 1.1 arcmin between difference scans.

To make a grid, we average 4-second spectra, which cover about 4 seconds
of integration time, corresponding to 60" in the RA direction. In this
way, we obtain a datacube with 1 arcmin spacing/gridding which is
slightly oversampled. We combine data from both the drift scanning and OTF mapping observations, and re-grid them with 1' spacing in the image plane using the method of \cite{Barnes2001MNRAS.322..486B}, 
and eventually create a datacube in the standard FITS format.

The rms is about 0.5 mJy per 5.1 km s$^{-1}$ channel, or 7.5mK in
Ta* in the final spectra. Assuming $\eta_A= 0.75$ (\citealt{XU2021ApJ...922...53X}), the corresponding column density sensitivity is $1\sigma = 9.8\times10^{16}$  
atoms cm$^{-2}$ per channel. The 5 $\sigma$ level is $4.9\times 10^{17}$atoms cm$^{-2}$.

\section{Results }


\begin{figure*}
   \centering
  \includegraphics[width=0.72\textwidth] {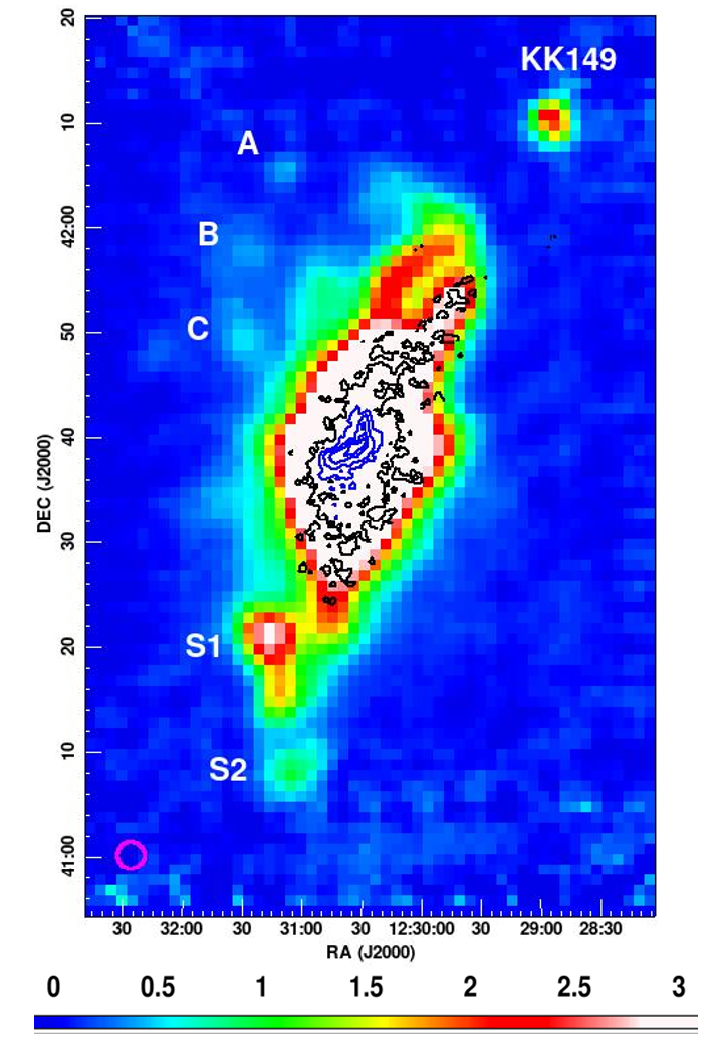}
   \caption{The HI integrated intensity map of
NGC 4490/85 ($395 < V_{HEL} < 726 km s^{-1}$) overlaid with contours of the WSRT interferometer map from
WHISP (\citealt{Van2001ASPC..240..451V})  with $30''$ resolution. The contour levels are 0.03, 0.18, 0.50, 1.0, 1.5 Jy/beam km s$^{-1}$, 
where the blue contours show the two disks of NGC 4485 and NGC 4490, and the black contours indicate the gas envelope detected by WSRT. 
The unit on the color map is Jy/beam km s$^{-1}$. 
 The pink  circle indicates the FAST beam size of $2.9'$.
   }
   \label{Fig.1}
   \end{figure*}

\begin{figure*}
   \centering
  \includegraphics[width=0.82\textwidth] {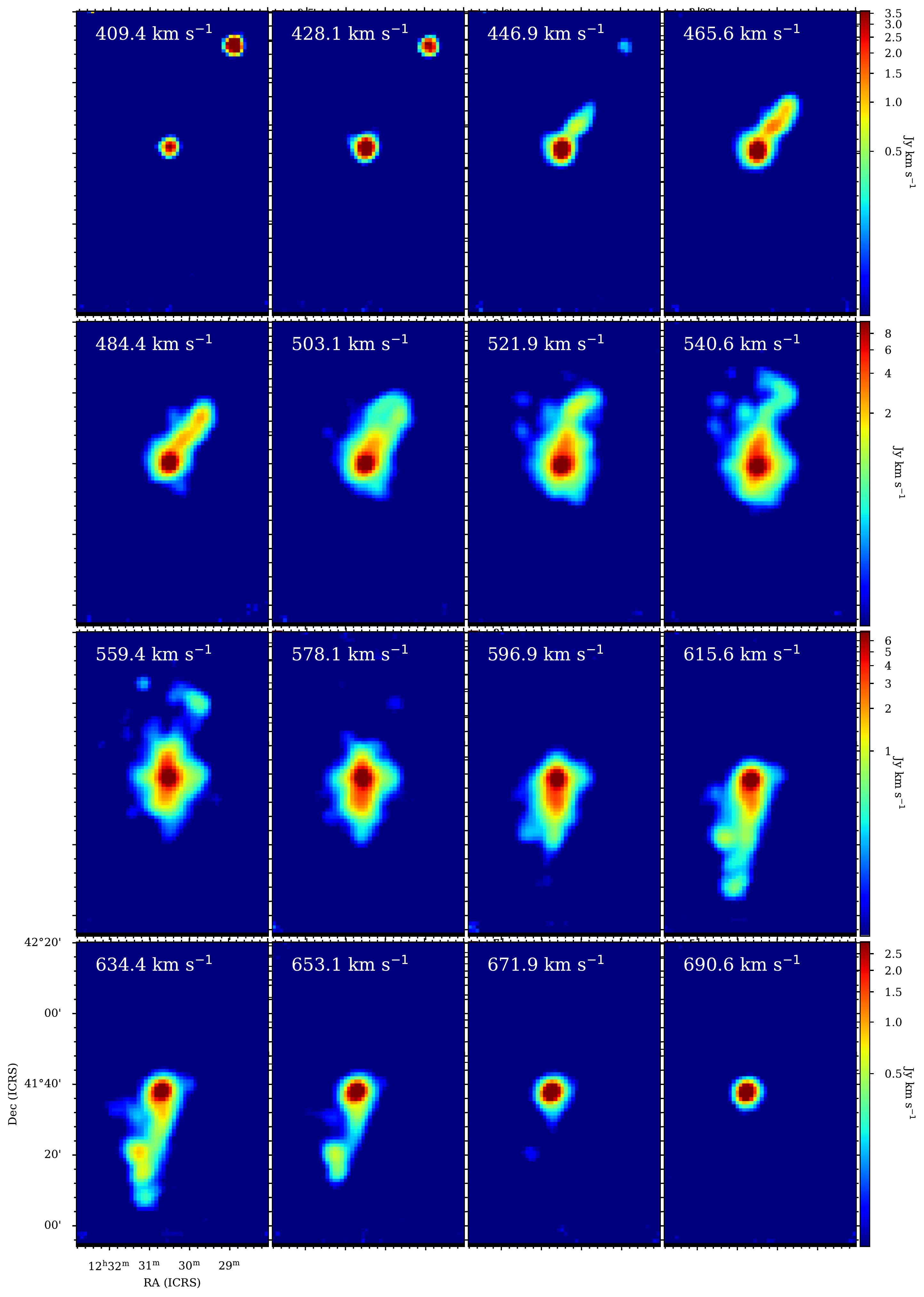}
   \caption{The channel map of the FAST HI data cube for NGC 4485/90. Each channel is integrated over a velocity range of 18.5 km s$^{-1}$.
   }
   \label{Fig.2}
   \end{figure*}

\begin{figure}
   \centering
  \includegraphics[height=9cm, angle=0] {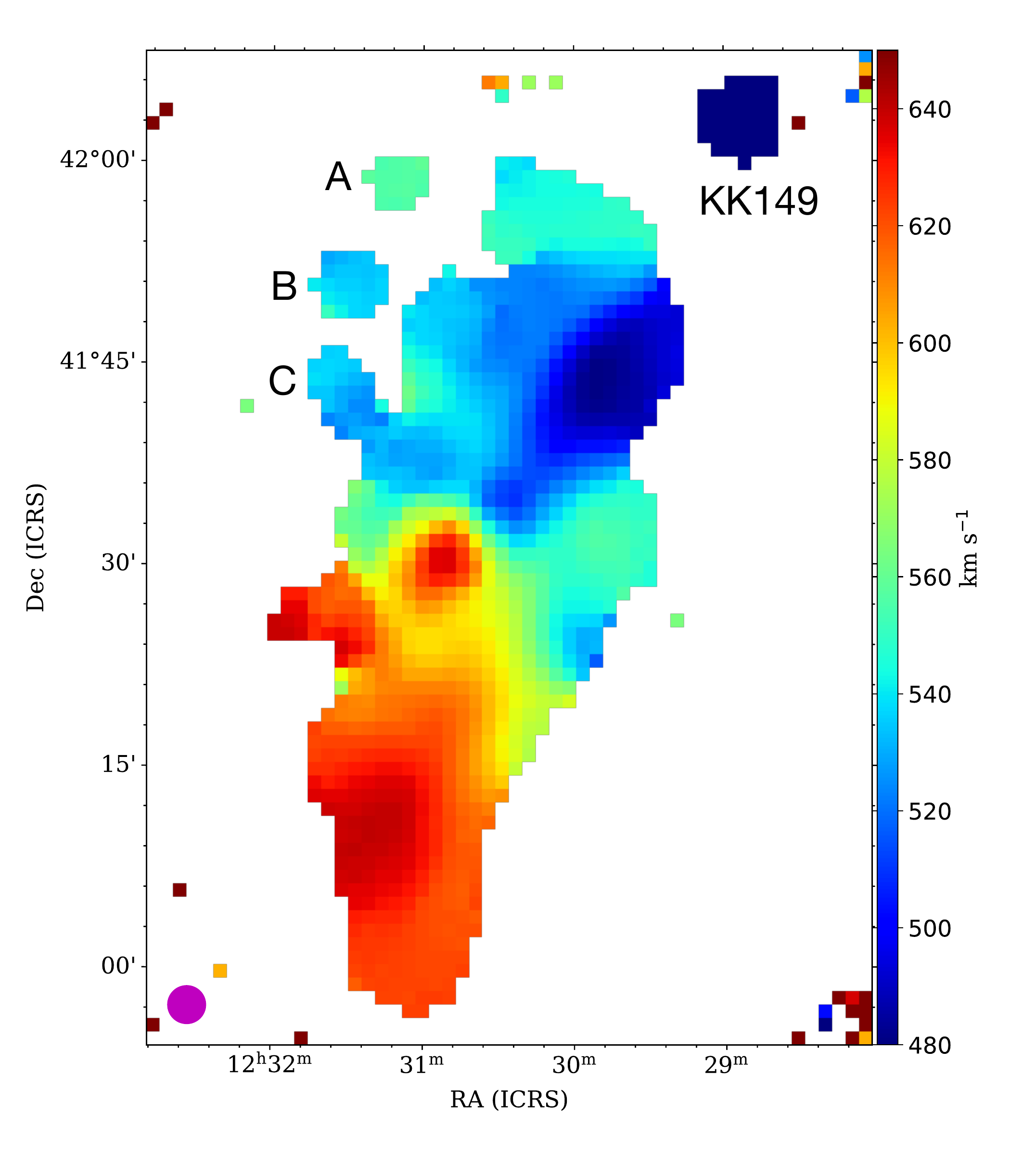}
   \caption{The momont-1 map of the FAST HI data cube for NGC 4485/90. The pink filled circle indicates a beam size of $2.9'$.
 }
   \label{Fig.3}
   \end{figure}

\begin{figure*}
   \centering
  \includegraphics[width=0.85\textwidth] {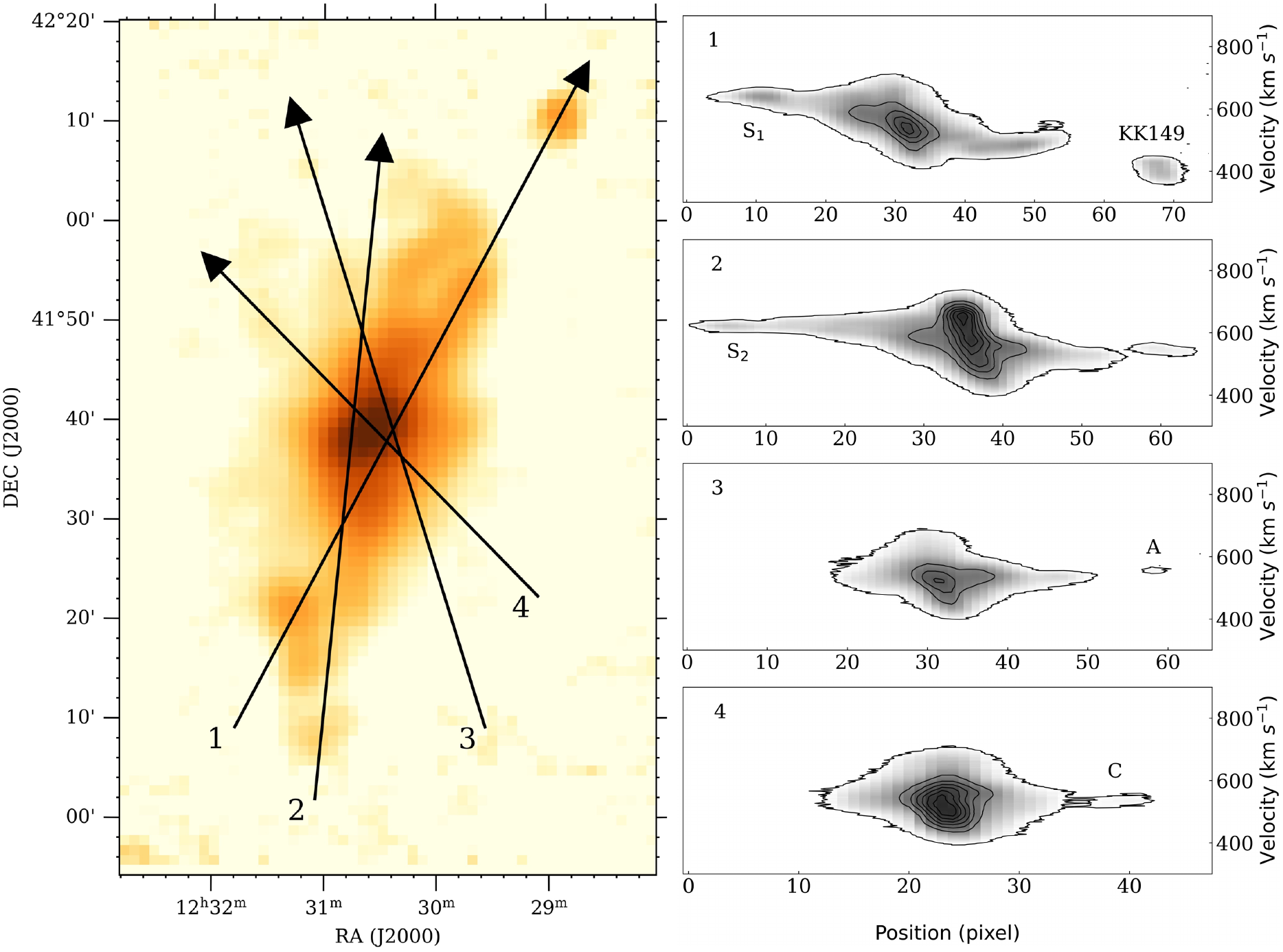}
   \caption{The position-velocity diagram  along 4 directions. Left:
The FAST H I integrated intensity map of NGC 4485/90. The four black arrows
mark the directions and positions of PV diagrams in panels 1-4, respectively.
Right: the black contours start at 4 $\sigma$ (3.8 mJy/beam) in steps of 15 mJy/beam.  Some 
H I features have been labbled.}
   \label{Fig.4}
   \end{figure*}

\begin{figure}
   \centering
  \includegraphics[width=0.85\columnwidth] {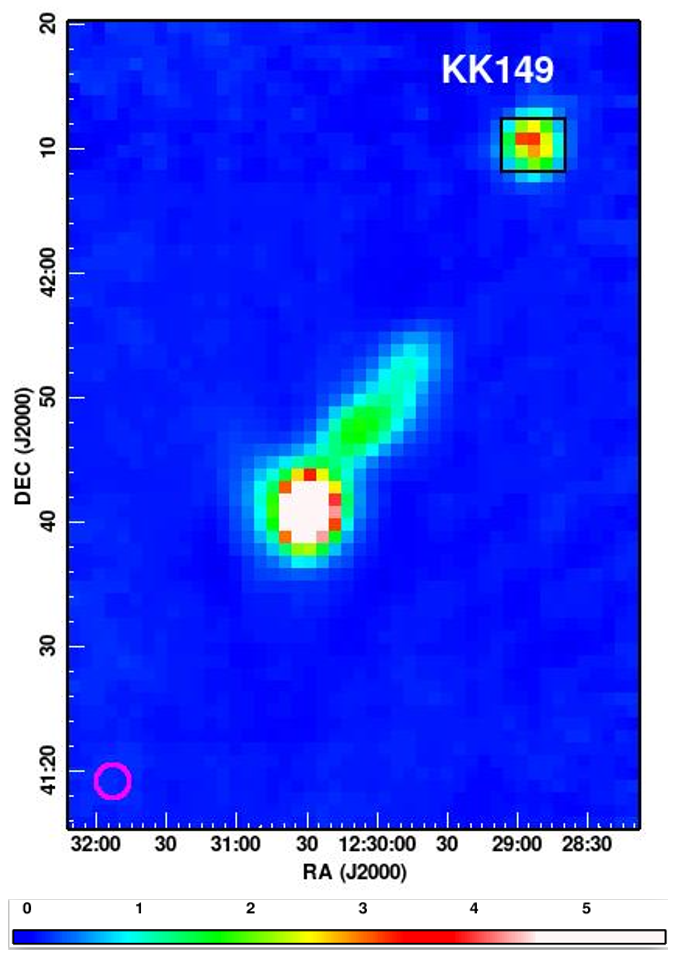}
   \caption{Left: The channel map of FAST HI data cube for NGC 4490/85. The HI fluxes are integrated over 349 to 470 km s$^{-1}$.  The pink  circle indicates the FAST beam size of $3'$. The black box indicates the region of the optical image at the right panel.The unit on the color map is Jy km s$^{-1}$. 
   }
   \label{Fig.5}
   \end{figure}


Figure~\ref{Fig.1} color map shows the FAST HI integrated intensity map of NGC 4490/85 ($395 < V_{HEL} < 725 km s^{-1}$) with a 3 sigma sensitivity of $4\times 10^{18}$ atoms cm$^{-2}$, while contours show the WSRT interferometer map from WHISP (\citealt{Van2001ASPC..240..451V}) for comparison.  The two galaxy disks are not resolved by FAST, but they can be seen clearly in the WSRT map.  The FAST map shows that the extended HI envelope surrounding the pair is much more extended (about 62'x26' in size) than that seen in the interferometer maps (40'x16').
This result is consistent with that of \cite{1980A&A....91..341H}, which estimated an HI envelope size of 57 x30 arcmin
(114x60 kpc), using the Effelsberg 100m radio telescope.
However, their resolution is 9.3', which is too low to show any detailed structures. The FAST map shows that the HI features can be better
described as a "HI envelope + tidal tails" structure. The envelope extended for about $40'$, and is symmetrically distributed on both sides of the NGC 4490 disk. The internal structure of the HI envelope was not clearly resolved by FAST. The detailed structure of the envelope can be seen in the VLA
map (\citealt{1998MNRAS.297.1015C}) or the WSRT map (\citealt{Van2001ASPC..240..451V}). However, both of these interferometric maps were limited by
the field of view and covered only the HI envelope. They did cover part of the northern tidal tail, and almost completely missed the southern tail. The FAST images reveal much more HI emission in a large region and recover much more fluxes than the interferometric maps.

The total flux of the NGC 4085/90 system is 478.1 Jy km s$^{-1}$,
while the WSRT and VLA found 291.95 (\citealt{Kovac2009MNRAS.400..743K}) 
and 305 Jy km s$^{-1}$ (\citealt{1998MNRAS.297.1015C}), respectively. We recovered 57\% more fluxes compared with the VLA map.  
Note that \cite{1980A&A....91..341H} got 525.86 Jy km s$^{-1}$ with 15\% uncertainty, using the
Effersberg 100m telescope and our result is consistent with theirs within the uncertainties.
The total HI mass for the whole system is $ 5.8\times 10^9 M_\odot$ computed as $M_{HI}=2.36\times  10^{5}\times  D^{2}\times f_{HI}$.
We further estimate the HI mass distribution in different parts of the system, which is listed in Table ~\ref{tab1}. The gas envelope contains about 46.7\% HI mass, and the two galaxy disks
contain 44\%. The southern and northern tail contains 4\% and 5\% HI mass, respectively.

\begin{table}
\centering
\caption{HI fluxes and masses for different parts of the NGC 4490/85 system}
\label{tab1}
\resizebox{\columnwidth}{!}{%
\begin{tabular}{ccccc}
\hline
Sources$^{a}$         & HI Velocity(optical) & HI fluxes        & lines width W50 & HI mass$^{b}$        \\
                & (km s$^{-1}$)        & (Jy km s$^{-1}$) & km s$^{-1}$     & M$_\odot$       \\ \hline
NGC4485/90 disk & -                    & 212.2            & -               & $2.6\times10^9$ \\
HI envelope     & -                    & 223.2            & -               & $2.7\times10^9$ \\
Southern tail   & -                    & 19.1             & -               & $2.3\times10^8$ \\
Northern tail   & -                    & 23.6             & -               & $2.8\times10^8$ \\
KK 149          & 413                  & 3.6              & -               & $4.6\times10^7$ \\
S1              & 638                  & 4.03             & 38              & $4.8\times10^7$ \\
S2              & 623                  & 1.02             & 18              & $1.2\times10^7$ \\
Clump A         & 556                  & 0.28             & 20              & $3.4\times10^6$ \\
Clump B         & 536                  & 0.75             & 32              & $9.0\times10^6$ \\
Clump C         & 537                  & 0.86             & 36              & $1.0\times10^6$ \\ \hline
\end{tabular}
}
\footnotesize{${^a}$ The NGC 4485/90 disk region is defined as the region inside the blue contour in Figure 1; the HI envelope is all the HI detected region excluding the two disks and the two tails. The southern and northern tails are defined as the tail region 
with Dec $<$ 41:26:06.5 and Dec $> $ 41:50:12.0, respectively; S1 and S2 are defined as  within a radius of $3'$ 
from the column density peak;
Clump A,B,C are defined as regions above 4$\sigma$. \\
$^{b}$ The HI mass is computed assuming all the sources at D=7.14 Mpc.}
\end{table}



Beside more fluxes detected, there are three new features in the FAST map, which are described in the following three sections.

\subsection{Longer tidal tails}

The tidal tails extended much further in the south and north directions. The Northern tail is 18', while the southern one is 31.5',
corresponding to 36 and 63 kpc, respectively. It is much longer than that seen in the WSRT map.
 Such asymmetry morphology is consistent with the simulation result of the tidal
interaction model by \cite{pearson2018modelling}.
Figure~\ref{Fig.2} shows the channel map of NGC 4485/90. 
The southern tail is visible in the range of 590-663 km s$^{-1}$ while the northern
extension range is 450-577 km s$^{-1}$.

Figure~\ref{Fig.3} is the moment-1 map from FAST, which shows the velocity
structure of the whole system. 
The southern tail appears to be from the high velocity part of the disk and 
the northern part occupies the low
velocity end which should be originated from the blue end of the disk.  These
features are also consistent with the tidal origin of the tails.

Figure ~\ref{Fig.4} shows the position-velocity (PV) diagrams, in which we extract four  slices
of the cube along different directions to illustrate the
 corresponding gas kinematics of features 
in NGC 4485/90.  Directions 1 and 2 are along the tidal tail directions and the HI feature shows a "HI envelope +
tidal tails" structure. The envelope has large velocity dispersion along all 4 directions.  
In directions 1, 3, and 4, we can also see a connection
between the gas envelope and KK 149, clump A, and clump C.

The total HI mass is $2.3\times 10^8$ M$_\odot$ in the southern
tail. There appear to be two clumps in the southern tail.  Clump S1
has an integrated flux of 4.03 Jy km s$^{-1}$ and its HI mass is about $4.8\times 10^7 M_\odot$,  similar to
that of a dwarf galaxy.  Clump S2 at the tail has a flux of 1.02 Jy km s$^{-1}$, corresponding to an
HI mass of $1.2\times 10^7 M_\odot$.  Such mass is high enough to form a tidal dwarf galaxy (TDG) in the future. From the Dark Energy Camera Legacy Survey(DECaLS) optical image, there are no optical counterparts associated with these two clumps. 
Thus they should be pure HI clouds.


\subsection{Complicated tidal structures in the northern tails}
The high sensitivity  FAST map reveals  detailed structures  of    
the northern tails which have three velocity components. These components can be better viewed in the channel maps in Figure~\ref{Fig.2}:

(1) One is a tail-like structure centering at V=485 km s$^{-1}$ and range from 450 km s$^{-1}$ to 520 km s$^{-1}$, namely the northern tail-1.

(2) A second tail-like structure starts to show up in the channel map
from v=503 km s$^{-1}$ and ends at V=551 km s$^{-1}$ with a central velocity of 530 km s$^{-1}$, namely the northern tail-2.

 
(3) From the channel map we can also see a third velocity component  appears at velocity 
channels of 522-559 km s$^{-1}$, 
centering at 540 km/s. This component contains 3 weak discrete clumps in the upper north of
N4485, distributed in a loop-like structure. 
Such features are also seen in Figure~\ref{Fig.1}.
We mark these clumps with A, B, C. Clump A coincides spatially and in velocity with a dwarf galaxy MAPS 1231+42
(MAPS-NGP O 218 0298413) at RA=12h31m09s, Dec=$42^{\circ}05'39"$, at the velocity channel of 563 km s$^{-1}$.  This galaxy is also detected
in the WSRT CVn survey (\# 43 in  \citealt{Kovac2009MNRAS.400..743K}). Comparing the fluxes we
measured, our flux value of 0.28 Jy km s$^{-1}$ (Table ~\ref{tab1}) is consistent with that of the WSRT result (0.21 Jy km s$^{-1}$).  
The most interesting point is that
\cite{2012A&A...546A.122I} found that this galaxy 
is extremely metal poor with O/H = 7.65,
which is only 5\% of the solar neighborhood value,
suggesting that  this galaxy might be a newly formed galaxy out of raw material. 
We will discuss its origin in more details in section 4.2.
We note that part of the HI tail forms a bridge connecting the tail with this
galaxy (see the channel map in Fig.3 with V=546-561 km s$^{-1}$), indicating that it could be due to tidal interaction with this galaxy. 

Clump B and C have a HI mass similar to that of Clump A, but are less compact. These three clumps may be formed out of the
HI loop which then fragments and collapses into individual clumps. Clump A
is more compact, which could be due to its connection with the dwarf galaxy.

For Clump A, the line width W$_{50}$ is 20 km s$^{-1}$  and W$_{20}$ is 36 km s$^{-1}$. The total 
HI mass is $3.4\times 10^{6}$ M$_\odot$. Based on the method described in \cite{GUO2020Nature}, we estimate that its dynamical mass $M_{dyn}=V_{HI}^{2}r_{HI}/G$  is about $3.8 \times 10^7$ M$_\odot$.

Table \ref{tab1} also list the HI fluxes and estimated mass of the three clumps. The origin of the loop-like structure is discussed in more details in the section 4.2.

\subsection{Connection with KK 149}

In Fig.1 and Fig.3, we can see a dwarf galaxy MCG+07-26-011 (KK 149) is also clearly detected at the position of RA=12h28m53.4s, Dec=$42^{\circ}10'37''$, at the velocity channel of 410km s$^{-1}$ (in Figure~\ref{Fig.1} and Figure~\ref{Fig.4}). Its velocity ranges from 361 to 451 km s$^{-1}$. 
 KK 149 is also detected
in the WSRT CVn survey (\# 44 in  \citealt{Kovac2009MNRAS.400..743K}). Comparing the fluxes we
measured, our flux value of 3.6 Jy km s$^{-1}$ is consistent with that of the WSRT result (3.48 Jy km s$^{-1}$), considering the uncertainty of about 10 percent.

Figure~\ref{Fig.5} shows the color map which is integrated over 349 to 470 km s$^{-1}$. This
map clearly shows a tail like structure pointing toward KK 149 (this structure is also seen on Figure~\ref{Fig.1}). Such good collimation suggests that NGC 4490 could be tidally interacting with KK 149 or there is a connection between them. The moment-1 map (Fig.3)
also shows that this tail has a similar velocity range as the northern part of NGC 4490 disk rotating at about 450km s$^{-1}$, thus this gas component could be from part of the NGC 4490 disk, which is dragged out of the disk and
moving toward KK 149.  This galaxy is also detected in the WSRT CVn survey (\citealt{Kovac2009MNRAS.400..743K}). The PV diagram from the WSRT data shows a screwed structure, with the high velocity end (above 450 km s$^{-1}$) being stripped away (see the Figure for WSRT-CVn-44 in Appendix A of \citealt{Kovac2009MNRAS.400..743K}). This part of the HI tail could also be from the gas
stripped from KK 149 when it penetrates through the gas envelope of NGC4485/90. We will discuss this scenario in more details in section 4.

The distance to KK149 is 8.51 Mpc (\citealt{tully2013cosmicflows}) based on TRGB
 measurements, which is similar to that of NGC 4485 in Tully's catalog of Cosmic flow 2.  It is about 50-60 kpc away from NGC 4490 in projection, and thus could be physically interacting with the N4490 pair.

\section {Discussion}
Our new HI data shows that the NGC 4490/85 pair has an ''envelope +tails'' structure.
They could have different origins. 

\subsection{ Origin of the tidal tails}

The tidal tail structures are one of the most common features seen in closely interacting pairs. Thus it is not a surprise to see two long
tails in NGC4490/85. Our FAST observations show that the velocity structure of the northern tails matches that of NGC 4485, thus it is very
likely that these tails are the debris torn out from the NGC 4485 disk due to interaction with NGC4490. \cite{pearson2018modelling} used N-body
simulations to model the interaction and they demonstrate that dwarf-dwarf interactions can form two long tidal tails
after 1.8 Gyr of encounter. Although this simulation was used originally to explain the gas envelope, our new data show that it can be used better to explain the tidal tail features, with an earlier tidal encounter time between the two galaxies (to have the gas to be more extended at the present day).  In fact, our new tail features resemble Pearson's model system about 0.5 Gyr into the future from the present
day at a slightly different viewing angle (see their Figure 3), i.e. the tidal tails revealed by FAST could be formed after 2.3 Gyr of encounter.

Although this model explains the tail morphology reasonably well, it also predicts that
the two disks should have merged into one disk by the time of 2.2 Gyr, which is not consistent with the configuration of NGC 4490/85. 
However, it should be possible to adjust the model parameters and eventually obtain a fully matched result.

On the other hand, \cite{lawrence2020revealing} found double nuclei in NGC 4490 using deep IR
imaging data. This suggests that  another merger event could have occurred, probably more than
2.3 Gyr ago. Thus the tidal tails could also be generated by this encounter event. It is possible that 
two encounter events acting together to produce the envelope and tidal tail features in NGC 4490/85.
In this case, a new model involving three closely
interacting objects are needed to explain all the features.

\subsection{Origin of the metal poor dwarf galaxy MAPS 1231+42}

As shown in Fig.1 and Fig.4, Clumps A, B, C appear to be in a loop-like structure, with the galaxy MAPS 1231+42 clearly 
associated with Clump A.
This galaxy is also detected by GALEX at UV band, with the name of GALEXMSC J123109.07+420538.1 and an FUV flux of $6.6\times 10^{-2}$ mJy, suggesting that star formation is actively on-going. 
\cite{karachentsev2013star} estimate that its SFR is -3.21 $M_\odot yr^{-1}$ in log scale(adjusted for distance). As its HI gas is $log M_{HI}=6.53 M_\odot$, the gas depletion time is about 
0.49 Hubble time. Thus the star formation in this galaxy could be triggered by the
 tidal interaction with NGC 4485/90  during the previous encounter. Its optical morphology
looks like a blue compact dwarf galaxy. 

One might think that this galaxy could be a tidal dwarf galaxy newly formed out of Clump A. 
However, what is surprising is that the galaxy MAPS 1231+42 is extremely metal poor.  
If it is a tidal dwarf galaxy that forms out of the pre-enriched disk tidal debris, its metallicity should be similar to that
of the NGC 4485/90. 
For comparison,  the metallicity of NGC4490 is $12+\log(O/H)=8.3 $ (\citealt{pilyugin2007oxygen}; \citealt{Taddia2013A&A...558A.143T}). 
We have also searched the SDSS DR 17 archive and found two spectra for NGC 4485. From these spectra, we can estimate that 
the gas-phase metallicities are in the range of $ 12+\log(O/H)=8.22-8.45$ in NGC 4485. 
The detailed calculations are presented in Appendix A. 

Thus the gas-phase metallicities of MAPS 1231+42 is much lower than that of the interacting 
pair NGC 4490/85, suggesting that the gas fueling the star-formation episode can only be
very metal poor and thus primordial, not out of the tidal debris of NGC 4490/85.

Another problem with the TDG scenario is that the time scale for the
formation of a new galaxy is too short.  The formation time scale for the NGC 4490/85 system is about 2.3 Gyr, and the gas envelope might not be well developed until about 1.8 Gyr based on the simulation of \cite{pearson2018modelling}. MAPS 1231+42 locates far away from the disk. 
It should be formed out of the gas envelope if it is a TDG.  Thus the age of the TDG should be less than about 0.5 Gyr old.  Although it is difficult to determine the age of MAPS 1231+42, due to the fact that the stellar light is dominated by a young population, this galaxy appears to have some old stars as it can be detected in r band or z band in the SDSS image. The age of these old stars should be older than 0.5 Gyr, longer than the TDG formation time scale.
Therefore, this galaxy is most likely an pre-existing dwarf galaxy instead of a tidal dwarf. 
In this case, the extremely low metallicity indicates that it should be in a dormant phase with virtually
no star formation (SF) before it is triggered by the NGC 4490/85 interaction. 
It could be entering an activated phase of SF due to gas accretion from the giant envelope of NGC 4490/85.

 What is puzzling is the formation of the loop-like structure containing Clumps A, B, and C.   
 This loop is far away ($23'$ ) from the NGC 4490 disk ($\sim $46 kpc), thus it should not be due to star
formation or supernova explosion on the disk.  It could be due to a head-on collision of the gas envelope with another galaxy, as collisional rings
are often found in the head-on galaxy collisions (\citealt{Appletion1996}; \citealt{Bekkj1998ApJ}; \citealt{Bournaud2003A&A}). One possibility is 
a collision with the galaxy
MAPS 1231+42, which has a dynamical mass of $3.8 \times 10^7 M_\odot$. 
Another possibility is a collision with the more massive galaxy KK 149 which is located in the northwest of the NGC 4490 pair. This galaxy is about $26'$ (55 kpc) from the
ring center. If it is moving at a speed of 100 km s$^{-1}$, the collision could
occur about 0.57Gyr ago, which is consistent with the formation time scale for the
ring. This galaxy has a $M_* \sim 2.6 \times 10^8 M_\odot$ 
(\citealt{Karachentsev2013b}) and
$M_{HI} =4.6\times10^7 M_\odot$  (D=7.14 Mpc), which should be able to produce part of the tidal
tail if it collides with the gas envelope. In section 3.3, we have provided two pieces of evidence to show that
KK 149 is interacting with NGC 4485/90. 

Besides direct collisions, other types of tidal interaction between KK 149 or MAPS 1231+42 and the gas envelope 
are also possible to generate the
chain of objects containing Clumps B and C and the loop-like structure. 
The N-body simulation of \cite{pearson2018modelling} did not take into account the interaction with
 small companions other than the NGC 4490/85 pair, and thus can not explain the formation of
the loop-like structure. More sophisticated modeling involving gas dynamics, star formations
gas outflow and accretion are 
needed to explain the formation of such type (loop-like) structure and model the 
evolution of the satellite galaxies MAPS 1231+42 and KK 149.

In summary,  the galaxy  MAPS 1231+42 is clearly associated with the tidal debris of the interacting pair.
Thus the intense star formation in it should be related to this interaction.

\begin{figure}
   \centering
  \includegraphics[width=\linewidth] {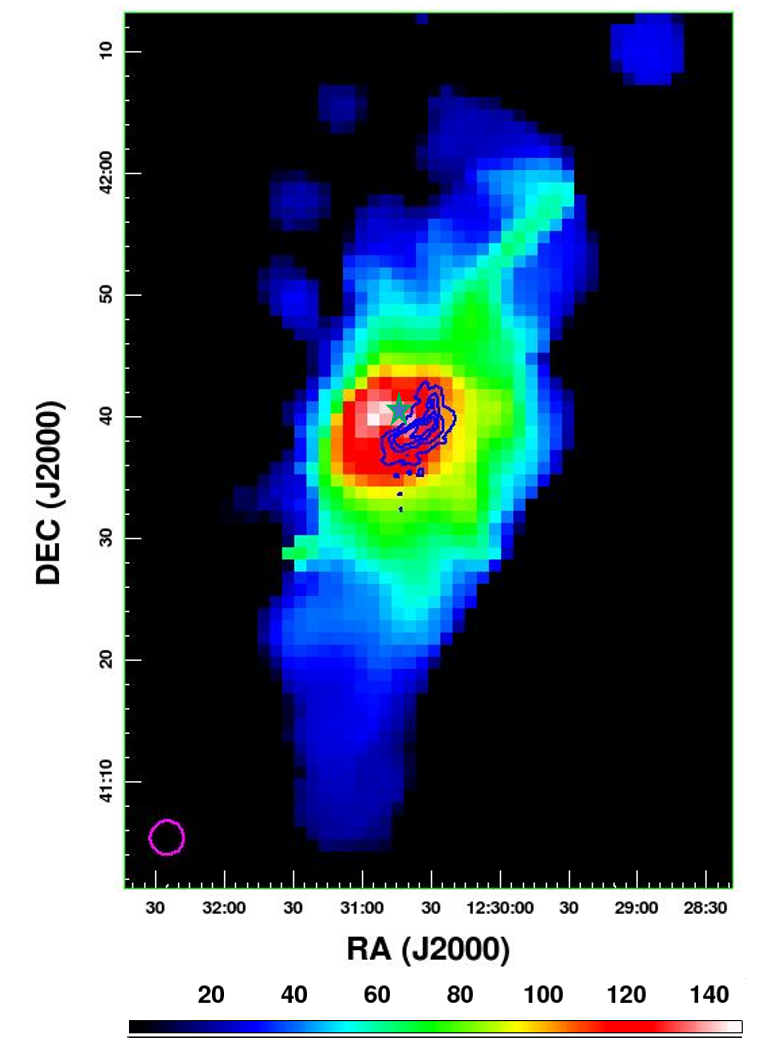}
   \caption{The momont-2 map of the FAST HI data cube for NGC 4485/90. The blue contours shows the two disks of NGC 4485 and NGC 4490 observed with WSRT.  The contour levels are 0.18, 0.50, 0.10, 1.5 Jy/beam  km s$^{-1}$. The green star indicates the location of an extended dust emission detected by the SCUBA of the JCMT. The unit for the color bar is km s$^{-1}$.
   }
   \label{Fig.6}
   \end{figure}

\subsection{Origin of the extended gas envelope}
NGC 4490/85 is well known for a long time for its giant gas envelope.
Based on the fact that the gas envelope is perpendicular to the NGC 4490 disk,
\cite{1998MNRAS.297.1015C} have shown that
 gas outflows from intensive star formation can provide a
simple interpretation for the origin of the envelope.  \cite{clemens1999star} further investigate the SF history in this
pair and find that the star formation in NGC 4490 has been on-going at about 4.7 M$_\odot$ yr$^{-1}$. 
This is consistent with the MIR 24um/FUV estimate of the SFR of about 3.4 M$_\odot$/yr by \cite{lawrence2020revealing}. 
Furthermore,  \cite{clemens1999star} estimated that the
duration of continuous high levels of star formation in NGC 4490
is about constant for $10^8$ yr. 
The \cite{Lotz2008MNRAS} models also found that the star formation
activity induced by the merger lasts for $ \sim 10^8 $ yr. 
These timescales are consistent with  the SF driven emplacement timescale of the NGC 4490 HI envelope ($6 \times 10^8$ yr, \citealt{1998MNRAS.297.1015C}).
Indeed, \cite{1998MNRAS.297.1015C} estimated
that on purely energetic grounds, only a few percent of the
mechanical energy output of each supernova is required.
\cite{Clemens2002MNRAS.333...39C} further pointed out that 
the atomic gas envelopes could act as large reservoirs to
steadily fuel star formation for over a periods of $10^9$ yr. 
Thus gas outflow from star formation should be able to explain the gas envelope.

 However, \cite{pearson2018modelling} argued that gas outflows are not needed to explain the extended gas envelope. 
They used N-boy simulations  to show that the large extended gases envelope 
morphology in the VLA map is actually the tidal tail structure viewing at a particular angle so that the tail is seen 
as a symmetric structure 
centered on NGC 4490.  Now that much longer tidal tails are observed by FAST, 
this model can not fit both the "tidal tail +envelope" features and can no longer be used to explain the gas envelope.

Our data also confirmed the large velocity dispersion found by \cite{1998MNRAS.297.1015C} that supports the gas outflow scenario.  Figure~\ref{Fig.6} shows the FWHM  map of the HI
datacube in the envelope.  It is clear that the velocity dispersion is
widely distributed all over the envelope region. The largest velocity
dispersion is located near the $ H_{\alpha}$ filaments found by (\citealt{Clemens2002MNRAS.333...39C}).  
An extended dust emission detected by the Submillimeter Common User Bolometer Array (SCUBA) of the JCMT, marked with a star in Fig~\ref{Fig.6}, is also found in this region 
(see Fig 4 of \citealt{Clemens2002MNRAS.333...39C}). 
However, given the relatively low spacial resolution of FAST, some of the wide spectral profiles could be due to more than one components superposed on it, as can be seen in the P-V diagram in Figure 4. Thus caution should be taken in interpreting the FAST FWHM map data.

Beside gas outflows, gas fallback from the tidal tails could also contribute to the gas envelope. 
However, both these two
scenarios would result in a metal-rich gas envelope, which is surprisingly in contradiction to
the extremely low metallicity  found in the dwarf galaxy
  MAPS 1231+42. This galaxy is experiencing a starburst phase.
  A straightforward speculation would be to consider the gas accretion
from the envelope as a trigger of the starburst. In such case, if a large amount of gas has been accreted, the metallicity should be similar to that 
of the envelope, thus our finding of such a low metallicity suggests that either we are capturing the early phase
of gas accretion and star formation in MAPS 1231+42, or part of the envelope gas surrounding MAPS 1231+42 is also
metal-poor and primordial.  

Nevertheless, beside gas accretion, other mechanisms are also possible to trigger the starburst in MAPS1231+42.
Detailed hydrodynamical simulations would be needed to model the interactions between MAPS 1231+42 and the NGC 4490/85 pair
together with the gas envelope to explain the star formation and metallicity in this dwarf galaxy.  While such theoretical work 
is beyond the scope of this paper, we like to point out that \cite{sparre2022gas}
model galaxy interactions based on cosmology simulation, they find
that gas accreted from the circumgalactic medium (CGM) during
the merger contributes a significant part (30-40 percent) to the star
formation as well as the gas bridge at the time of coalescence.
Such simulation results could explain the low metallicity in MAPS
1231+42, as part of the gas it accretes could be originally from the
metal-poor CGM environment.   
Future metallicity measurements of the envelope gas are needed to confirm or reject this scenario. 
In fact, \cite{Wang2023ApJ} have  found a large amount of diffuse HI gas surrounding the 
interacting pair NGC 4631/56, and they suggest that the diffuse HI in the tidal tail region 
can induce cooling out of the hot intragroup medium.


Extended gas envelopes are also seen in other starburst galaxy such as IC10 (e.g. \citealt{Nidever2013}), 
which is not in clear interacting pair. Another example is NGC 4449 (e.g. \citealt{Bajaja1994}, \citealt{Hunter1998}), 
for which only weak interactions are seen.  We expect that more cases of extended  gas envelopes will be found with high sensitive observations using the large single dish telescope FAST.

\section{Summary and conclusion}
With FAST's superior sensitivity, we discover a much more diffuse HI structures in the interacting galaxy pair NGC 4490/85. Our finds are summarized as the following: 

1) We found the tidal tails in NGC 4490/85 are much longer than that seen in the VLA interferometric map. 
The total flux measured by FAST is 40\% more than that measured by the VLA. 
 The general HI distribution has a "tidal tails+envelope"  morphology, and 
the N-body simulations of \cite{pearson2018modelling} which was designed to match the VLA map 
can not reproduce all the features found by FAST. A similar model with a modified viewing angle and 
encounter parameters should be able to reproduce the long tidal tail features. 
We  confirm the large velocity dispersion in the gas envelope found in the VLA map, supporting the gas
outflow origin proposed by \cite{clemens1999star}. 
Hydrodynamical simulations involving gas dynamics
e.g. gas outflows from star formation feedbacks would be able to explain the formation of the gas envelope.

2) We found a collimated gas component pointing at a nearby dwarf galaxy KK 149, suggesting that this galaxy  might also be interacting with the NGC 4490/85 pair.

3) We found several tidal features in the gas envelope of NGC 4490/85 which 
is connecting with a starburst low metallicity dwarf galaxy  MAPS 1231+42, suggesting that galaxy 
interaction have triggered  intensive star formation in it. 
 This result shows that dwarf interacting pair can also have impacts on the evolution of 
other dwarf galaxies in its vicinity.  Such small companions and their tidal features, 
with a column density of about $1\times 10^{19}$ cm$^{-2}$, 
can only be detected in the local universe even with telescopes as big as FAST.  
Many distant interacting systems may also have interactions with small companions but 
might have been missed by the current generation of radio telescopes due to limited sensitivity.
Therefore, galaxy interactions may be more common than we thought, even in dwarf-interacting pairs. 
Future deep observations with the next generation of radio telescopes such as the SKA should be able to detect more weakly
interacting systems.



\section*{Acknowledgements} 
We thank the anonymous referee for insightful comments and
constructive suggestions.  We acknowledge the support of the National Key R$\&$D Program of China (2018YFE0202900; 2017YFA0402600). This research made use of data from WSRT WHISP. The Westerbork Synthesis Radio Telescope is operated by ASTRON (Netherlands Institute for Radio Astronomy) supported by the Netherlands Foundation for Scientific Research NWO.  The Five-hundred-meter Aperture Spherical radio Telescope (FAST) is a Chinese national mega-science facility, funded by the National Development and Reform Commission. FAST is operated and managed by the National Astronomical Observatories, Chinese Academy of Sciences.

\section*{Data Availability}
The raw data used in the article will be published on the FAST website: \href{https://fast.bao.ac.cn}{https://fast.bao.ac.cn}.
The PID is N2021$\_$4.
Please contact the author (liuyao@nao.cas.cn, mz@nao.cas.cn) for processed data.


\appendix
\section{Estimate the metallicity in NGC 4485 }

\begin{figure}
   \centering
  \includegraphics[height=6cm, angle=0] {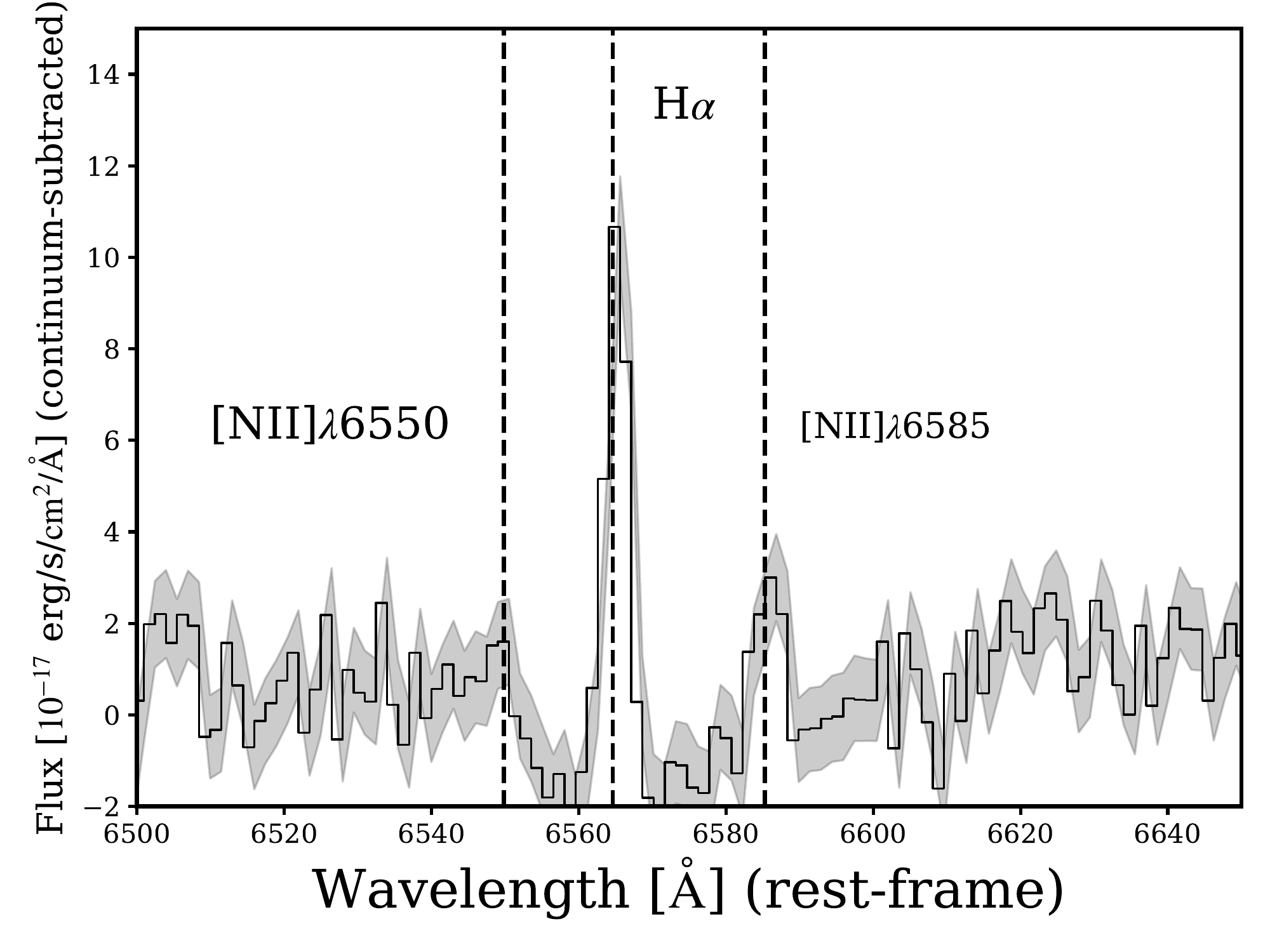}
   \caption{The SDSS spectum at the nuclear region of NGC4485 (RA=12:30:31.56, DEC=+41:42:01.0).
   }
   \label{Fig.a1}
   \end{figure}

\begin{figure}
   \centering
  \includegraphics[height=6cm, angle=0] {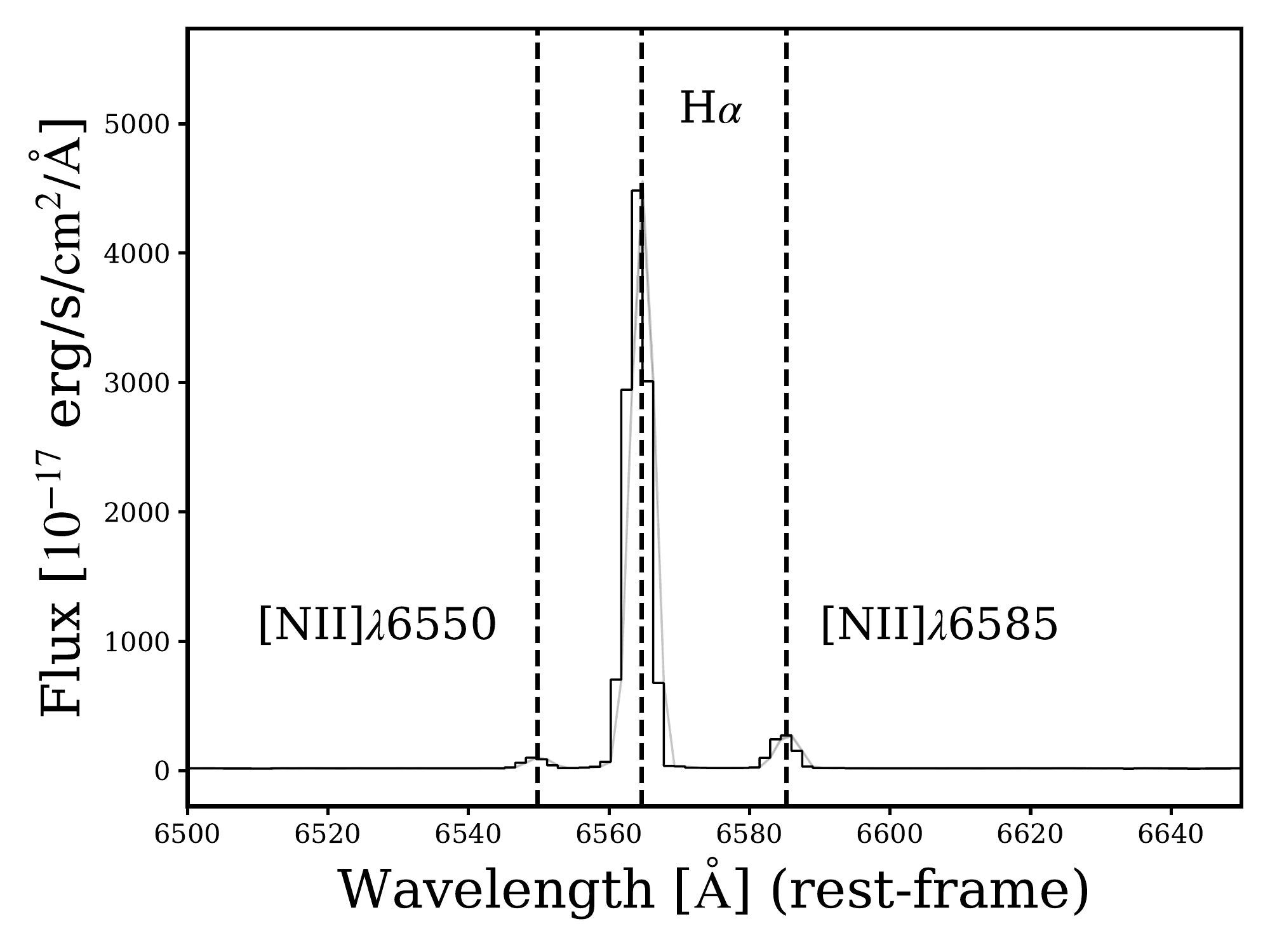}
   \caption{The SDSS spectum at an HII region near the bridge between NGC4485 and NGC 4490 (RA=12:30:28.33, DEC=+41:41:22.1).
   }
   \label{Fig.a2}
   \end{figure}

Checking the SDSS DR 17 archive, 
we found two spectra for NGC 4485,  one at the center of this galaxy (sources ID 1454-53090-316, Figure~\ref{Fig.a1}) and 
the other (source ID  1452-53112-11, Figure~\ref{Fig.a2}) at an HII region  near the NGC 4485/NGC4490 bridge. 

Using the N2-based method \citep{Van1998AJ,Pettini2004MNRAS,Marino2013A&A559A.114M}, we 
empirically calculate the gas-phase metallicity for these two regions.
The definition for this index is 
\begin{eqnarray}
    \rm N2 = \rm \log_{10}([NII]\lambda6585/H\alpha),
\end{eqnarray}
and this index can be converted to gas-phase metallicity with the following relation:
\begin{eqnarray}
\rm 12+\log(O/H) = \rm 8.90 + 0.57 \times N2.
\end{eqnarray}
The gas-phase metallicities are $\rm 12+\log(O/H)=8.45\pm0.15$ for source 1454-53090-316 and $8.22 \pm 0.007$ 
for source 1452-53112-11.

\bsp
\label{lastpage}
\end{document}